\documentclass{elsart}
\usepackage{graphicx}
\usepackage{amssymb}
\begin{document}
\begin{frontmatter}
\title
{ \Large \bf Role of connectivity in congestion and decongestion 
in networks
}

\author{
Neelima Gupte $^1$  and Brajendra K. Singh$^{2,*}$}
\address{ $^1$Dept. of Physics,\\
Indian Institute of Technology, Madras,\\
Chennai, 600036, INDIA.\\
$^2$ Institute of Physics, \\
Academia Sinica, \\
Taipei 11529, Taiwan.\\}
\begin{abstract}
{
We study network traffic dynamics in a two dimensional communication
network
with regular nodes and hubs.
If the network experiences heavy message traffic, congestion occurs due
to finite
capacity of the nodes. We discuss strategies to manipulate hub
capacity and hub
connections to relieve congestion and define a coefficient of
betweenness centrality
(CBC), a direct measure of network traffic, which is useful for
identifying hubs which
are most likely to cause congestion. The addition of assortative
connections to hubs
of high CBC relieves congestion very efficiently.
}
\end{abstract}

\end{frontmatter}
\section {Introduction}
Most communication networks seen in every day life suffer congestion 
problems at times of peak traffic. Telephone networks, traffic networks
and
computer networks all experience serious delays in the transfer of
information due to congestion or jamming\cite{Jamming}. Network congestion
occurs when too many hosts simultaneously try to send too much data
through a network.
Various factors such as capacity, band-width and network topology play
an important role in contributing to traffic congestion. The
identification  of optimal structures that minimise congestion as 
well as the identification of processes that give rise to such
structures have been considered in recent
studies\cite{Optstruc,Gradient} . However, there have
not been many attempts to improve the performance of
communication networks by making small modifications to existing
networks.
It has been established that the manipulation
of node-capacity and network
capacity can effect drastic improvement in the performance and
efficiency
of load-bearing networks \cite{Janaki}. Protocols which
can efficiently manipulate these factors to relieve congestion at high
traffic densities in communication networks can be of practical
importance. In this paper, we discuss efficient methods by which traffic
congestion can be reduced in a two dimensional
communication network of hubs and nodes 
by minimal
manipulation of its  hub capacities and connections
\cite{Braj,Braj1}. We set up a coefficient of betweenness centrality (CBC), which is a
direct
measure of message traffic \cite{BC},
and conclude
that the addition of assortative connections to the hubs of the highest
CBC                                                           
is the most effective way to                                  
relieve congestion problems.

\begin{figure}[tbp]                                           
\begin{center}
\includegraphics[scale=0.5]{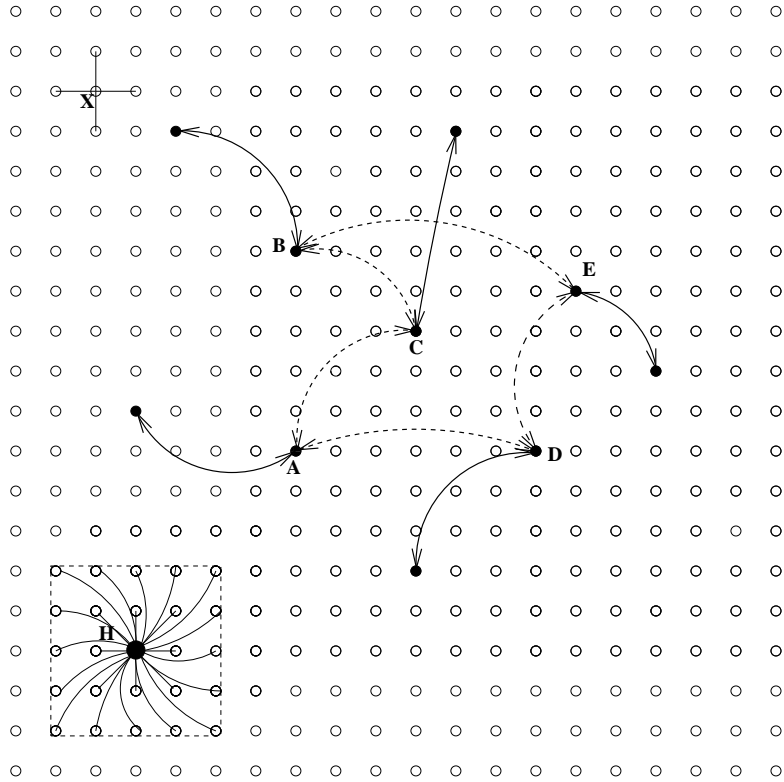}
\caption{\label{fig:epsart} A  2-d lattice with               
regular nodes with $4$ edges connected to all nearest neighbours (as        
explicitly shown for node $X$) and                            
hubs (filled circles) connected to  all constituent nodes within their      
influence area (see the hub $H$).
Two way assortative connections of two types are shown 
i) with dotted lines between any two of the top five hubs (labelled       
$A-E$), ii) with solid lines when the other end point     
is selected randomly from the rest of the hubs.}  
\end{center}            
\end{figure}    

\section{The model and routing}

We study traffic congestion for a
model network with local clustering\cite{Braj,Braj1}.          
This network consists of a two{-}dimensional lattice with two types of
nodes,
ordinary nodes and hubs (See Fig. 1).  Each ordinary node     
is connected to its nearest-neighbours, whereas               
the hubs are connected to                                     
all nodes  within                                             
a given area of influence                                    
defined as a square of side $2k$ centered around the hub\cite{FN}.          
The hubs are randomly distributed on the lattice              
such that no two hubs are separated by less than a minimum distance,        
$d_{min}$. Constituent nodes in the overlap areas of hubs  acquire
connections to all the hubs whose influence areas overlap.    
The degree distribution of this network is bi-modal in nature.    
There are several studies which examine                       
traffic on two-dimensional regular lattices\cite{Kleinberg}
as well as on                                                   
two-dimensional regular lattices with nodes of two            
types, which are                                              
designated as hosts and routers \cite{2d}.
It has been established that despite the regular geometry, traffic on       
such networks reproduces the characteristics of realistic internet
traffic. 

We simulate message traffic  on this system. Any              
node can function as a source or target node for a            
message and can also be a temporary message holder or router. 
The metric distance between any pair of source                
($is,js$) and target ($it,jt$) nodes on the network is        
defined to be the Manhattan distance $D_{st}=|is-it|+|js-jt|$.
The traffic flow on the                                       
network is implemented using the following routing algorithm.

Since the shortest paths between source and
target pairs on the lattice go through hubs
messages are routed through hubs.
 The current message holder 
$i_t$ tries to send the message towards  a temporary
target $H_T$, which is 
  the hub nearest $i_t$ which is closer to the target than
 $i_t$. If $i_t$ is an ordinary node, it send sends the message to its 
 to its nearest
neighbour towards $H_T$, or                                              
if  $i_t$ is a hub, it forwards the message to that of its
constituent nodes nearest to the final target.      
If the would-be recipient node is occupied, then the message waits      
for a unit time step at  $i_t$. If the desired node is still  
occupied after the waiting time is over, $i_t$ selects        
any unoccupied node of its remaining neighbours and hands over the
message.

In case all the remaining neighbours are occupied, the        
message waits at $i_t$ until one of them is free.             
When a constituent node of $H_T$, receives the message, it 
sends the message directly to the hub. If $H_T$ is occupied, then the      
message waits at the constituent node until the hub is free.  
When the message reaches the temporary target  $H_T$ it sends 
the message to a peripheral node in the direction of the target, which
then
chooses a new hub as the new temporary target and sends a message in its    
direction.         

\section{Congestion and decongestion:}

Although the hubs provide short paths on the lattice, hubs which have
many paths running through them  also 
function as trapping sites for messages due to their finite capacity. 
Such hubs can be identified using a quantity, 
the co-efficient of  betweenness centrality (CBC), which is a direct
measure of network traffic and 
defined as the ratio of the number of messages $N_k$ which pass through
a given hub $k$ to the total number of messages
which run simultaneously, i.e.
$CBC=\frac{N_k}{N}$.

\begin{figure}
\begin{center}
\includegraphics[scale=1.0]{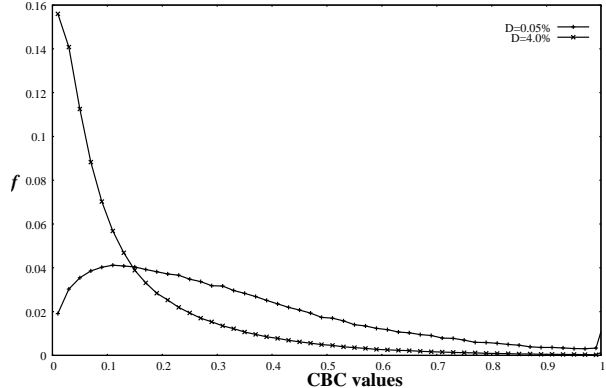}
\caption{\label{fig:epsart} This figure plots the fraction of hubs with
a given value of $CBC$  versus $CBC$
 values for $D=0.05\%$ and $D=4.0\%$.}
\label{fewhubs}
\end{center}
\end{figure}

We plot the distribution of the fraction of hubs with a given value of
$CBC$ against $CBC$ in Fig. \ref{fewhubs}. It is clear that 
hubs with low values of  $CBC$
dominate the distribution,
and 
the number
of hubs with high values of $CBC$ is very small. These hubs tend to
be potential locations of congestion. 
Additionally, the behaviour of many
communication networks in real life also indicates that a few hubs may be
responsible for the worst cases of congestion, and the significant
addition
of capacity at
these hubs alone may go a long way towards relieving network
congestion.
In order to test this idea, we operate our network in a regime where
congestion is likely to occur.                                                              

We compare the performance of the enhancement methods outlined above for
a network of ($100\times 100$)
nodes with overlap parameter $d_{min}=1$ for hub densities upto $4.0\%$.
The total number                                                     
of messages $N_m =2000$ and  $D_{st}=142$. The  length of the run is
fixed at $4 D_{st}$.
The average fraction of messages which reach their destination and the
average
travel time of the messages which reach are measures of the efficiency
of   
the network and are calculated over $1000$ configurations.
We test the baseline network, where each hub has unit capacity and can
only hold a maximum of one message at a given time, for its efficiency in terms of the number
of messages delivered as a function of the hub density. 
Table I lists the fraction of messages which reach their target as a
function of hub density. The hub density is listed in column one of the
table and the fraction of messages which reach the target for the
baseline in column two. It is clear that at low hub densities barely 
$6$ percent of the messages reach the target. 

To check whether the augmentation of capacity at the hubs of high
betweenness centrality relieves the congestion,  we augment the capacity of the top five hubs ranked by
their CBC by a factor of five (each of the top five hubs can now hold
five messages at a time).
Column three shows the
fraction of messages which reach the target for this enhanced case. 
Unfortunately, the comparison of the second and third columns indicates
that the capacity enhancement enhances the fraction of messages
delivered only marginally. Thus the enhancement of capacity alone
does not relieve congestion very significantly.

\begin{table}
\caption{ This table shows $F$ the  fraction of messages delivered
during a run as a function of the hub density $D$. The second  column
shows $F$ for the baseline viz. the lattice with hubs of unit capacity
and the remaining columns show the  fraction of messages delivered for
the case with enhanced capacity $CBC$, and the case of enhanced capacity
with assortative connections between the top five hubs ($CBC_A$) and
between the top five hubs and randomly chosen other hubs ($CBC_B$).}

\begin{center}
\begin{tabular}{|c|c|c|c|c|}
\hline
  D & $F_{Base}$ &  $F_{CBC}$ & $F_{CBC_{A}}$ & $F_{CBC_{B}}$\\ 
\hline
  0.10&      0.06225&          0.18260& 0.66554 &  0.75690 \\  
  0.50&      0.17441&          0.27144& 0.58882 &  0.70206 \\
  1.00&      0.30815&          0.39229& 0.72041 &  0.81193 \\
  2.00&      0.51809&          0.60946& 0.88792 &  0.92364 \\ 
  3.00&      0.68611&          0.77793& 0.95901 &  0.96914 \\
  4.00&      0.81786&          0.89181& 0.98536 &  0.98860 \\    
\hline
\end{tabular}
\end{center}

\end{table}

Earlier studies on branching hierarchical networks indicate that the
manipulation of capacity and connectivity  together can lead to major       
improvements in the performance and efficiency of the network 
\cite{Janaki}.                                                
In addition, studies of the present  network \cite{Braj}  
indicate that the introduction of  a small number of assortative
connection per hub  has a                                     
drastic effect on the travel times of messages. It is therefore
interesting to investigate whether introducing                
connections between hubs of high $CBC$ has any effect on relieving
congestion. We therefore add two way connections between the top
five hubs with enhanced capacities as above ($CBC_{A}$).
 The fraction of messages delivered is listed in the fourth 
column of table I. It is clear that there is a dramatic enhancement in
the number of messages delivered going from $6 \%$ to $66 \%$ at low hub
densities.                                             
Setting up  two-way connections between the top $5$ hubs
and randomly chosen other hubs ($CBC_{B}$) increased the number of messages which
were successfully delivered  to $75 \%$ (see the fifth  column of the
table). Thus the addition of assortative connections to a few hubs 
of high capacity relieves congestion very efficiently.

\begin{figure}                                                
\begin{center}
\includegraphics[scale=1.0]{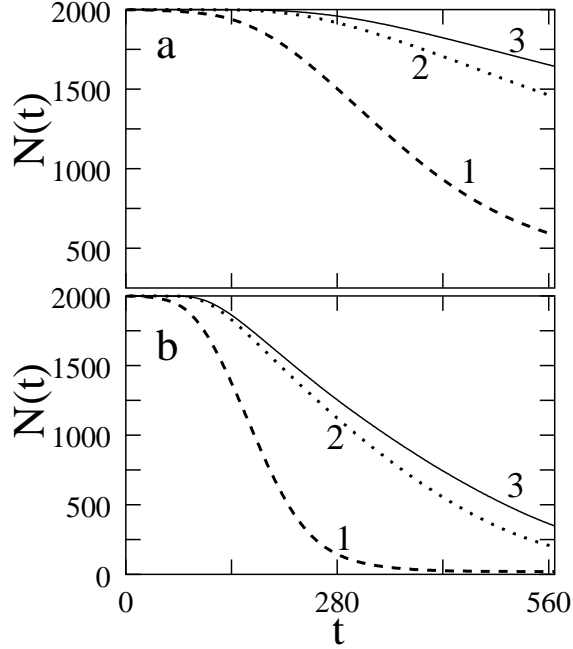}                
\caption{\label{fig:epsart} Plot of $N(t)$, the number of messages
running on the lattice as a function of $t$ at (a) low hub density (50
hubs), (b) high hub density (400 hubs). The curve labelled `1' shows
the behaviour on the lattice with assortative connections, the curve
labelled `2' shows that of the lattice with enhanced capacity ($CBC_2$)
and that labelled `3' shows the behaviour of the base-line.}
\label{Navg}
\end{center}
\end{figure}

The quantity $N(t)$, the total number of messages running in the system
at a given time $t$, is also a useful quantifier of the efficiency of
the   
system in delivering messages, as the number of messages decreases 
as they are delivered to the desired target. We plot this quantity in
Fig.~\ref{Navg}(a) (low hub densities) and Fig.~\ref{Navg}(b) (high hub densities) for the
four cases defined above. It is clear that the
addition of  two-way connections from the top five            
hubs (after capacity augmentation) to randomly chosen hubs from the
remaining hubs relieves the congestion extremely rapidly in comparison
to the base-line at both low and high hub densities.
 
\subsection{Queue lengths}

\begin{figure} 
\begin{center}
\includegraphics[scale=1.0]{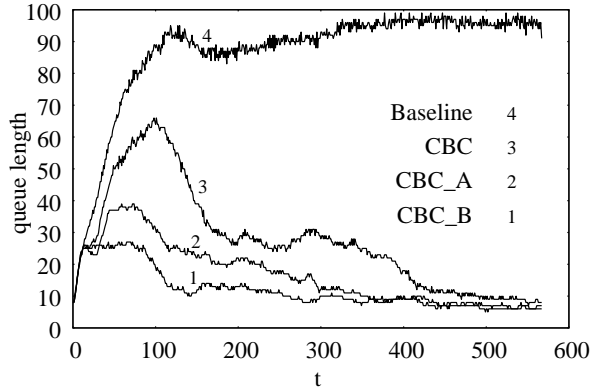}
\end{center}
\caption{\label{fig:epsart} The behaviour of queue lengths at the hub
with the  highest $CBC$ as a function of time.
The hub density was fixed at $0.05\%$.}
\label{qlength}
\end{figure}

Another interesting quantity in this context is the queue length at a given
hub as a function of time. A hub is said to have a queue of $N$ messages
if
at a given time $t$ all $N$ messages have chosen this hub as their temporary
target
during their journeys towards their respective final targets.
Fig.~\ref{qlength} shows the queue lengths as functions of time for one
of the top five
hubs for the base-line, CBC and the two cases of CBC with assortative
connections. It is very clear that the assortative connections clear the
queues very fast at each of the hubs by diverting messages along other 
paths. The queue lengths at several of the hubs show a peak before they
start falling off, indicating that the messages start taking alternate
paths only after the congestion along the shortest paths builds up.
\subsection{Average waiting times at constituent nodes}

\begin{figure}
\begin{center} 
\includegraphics[scale=1.0]{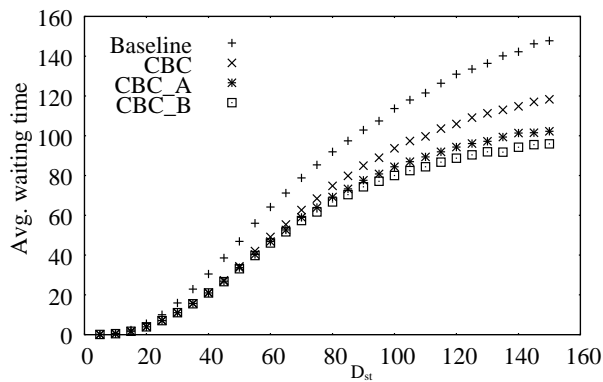}
\end{center}
\caption{\label{fig:epsart} Plot of average waiting time per message
as a function of $D_{st}$ for $N_m=1000$.
The hub density was fixed at $0.05\%$.}
\label{figphase1}
\end{figure}

We next look at the statistics of average waiting times. According to
our routing rules, a message waits at  the constituent node of a hub if
the delivery of the the message to the hub will exceed the capacity of
the hub.  
Thus the average waiting time, viz. the amount of time, on average, that
a message spends waiting at all the constituent nodes it encounters
during its entire journey, is an important characteriser of transmission
efficiency. We study the waiting time as a function of $D_{st}$ for the
different strategies. We also include the waiting time of messages which
do not succeed in reaching the target in this average. 
When most messages get through, this quantity has a small value (as at
low values of $D_{st}$ in Fig.~\ref{figphase1}) but it  increases in a
nonlinear fashion with increasing distance. The decrease in waiting
times  of the $CBC$
and $CBC$ with assortative connections when compared with the base-line
is clear from the figure.

\section{Discussion}

We thus see that  the addition of assortative connections to hubs of
high betweenness centrality is an extremely efficient way of relieving
traffic congestion  in a communication network. While the augmentation of capacity at such
hubs can  reduce congestion marginally, the data indicates that a large  augmentation of capacity
would be required to achieve effective decongestion. Thus the cost 
of achieving decongestion by capacity augmentation alone would be quite
high.
On the other hand, efficient
decongestion can be achieved by the addition of extra 
connections to a very small number of hubs of high betweenness
centrality. Decongestion is achieved most rapidly when two-way
connections are added from the hubs of high betweenness centrality to
other randomly chosen hubs. However, other ways of adding assortative
connections such as one way connections, or one-way and two way
connections between the hubs of high CBC also work reasonably well.
We note that this method is a low cost method as very few  extra
connections are added to as few as five hubs. 
The methods used here are general and can be carried over to other types
of networks as well.
Thus, 
our methods could find useful applications in realistic situations.
Our network with assortative connections is an example of an engineered network.
It would be interesting to see whether networks can
develop such assortative connections by self-organisation mechanisms.
We hope to report on these questions in future work.

\section{Acknowledgment}

NG thanks BRNS, India for partial support. BKS thanks BRNS, India, and
NSC, Taiwan, for partial support.

\vskip0.5cm

\small{$^*$Present address:
Dept. of Infectious Disease Epidemiology\\
St. Mary's Campus, Imperial College,
London W2 1PG, U.K.\\}


\begin{thebibliography}{99}
\bibitem{FN}                                                  
The shape of the influence area does not affect the results seen
\cite{Braj}.
\end{thebibliography}
\end{document}